\documentclass[times, twoside]{Preprint}

\usepackage{blindtext}
\usepackage{graphicx}
\usepackage{dblfloatfix}
\usepackage{lineno}
\usepackage{comment}
\usepackage[normalem]{ulem}
\usepackage{color, colortbl}
\usepackage{MnSymbol}
\definecolor{salmon1}{rgb}{1.0, 0.63, 0.48}
\definecolor{salmon2}{rgb}{1.0, 0.87, 0.68}
\definecolor{salmon3}{rgb}{1.0, 0.94, 0.84}
\definecolor{gray1}{rgb}{0.7, 0.7, 0.7}
\definecolor{gray2}{rgb}{0.85, 0.85, 0.85}
\definecolor{gray3}{rgb}{0.95, 0.95, 0.95}
\definecolor{blue1}{rgb}{0.69, 0.77, 0.87}
\definecolor{blue2}{rgb}{0.94, 0.97, 1.0}
\definecolor{blue3}{rgb}{0.88, 1.0, 1.0}

\leadauthor{M. Ballester} 

\begin{document}

\setlength{\parindent}{1cm}

\title{Modeling and Simulation of Charge-Induced Signals in Photon-Counting CZT Detectors for Medical Imaging Applications}

\shorttitle{Main Paper}

\setlength{\affilsep}{7 pt}
\renewcommand\Authfont{\color{black}\normalfont\sffamily\bfseries\fontsize{9}{11}\selectfont}
\renewcommand\Affilfont{\color{black}\normalfont\sffamily\fontsize{6.5}{8}\selectfont}
\renewcommand\Authands{, and }

\author[1]{Manuel Ballester}
\author[2]{Jaromir Kaspar}
\author[2]{Francesc Massan\'es}
\author[3]{Srutarshi Banerjee}
\author[2]{\\Alexander Hans Vija}
\author[1, 4]{Aggelos K. Katsaggelos}

\affil[1]{Department of Computer Sciences, Northwestern University, Evanston, IL 60208, USA}
\affil[2]{Siemens Medical Solutions USA Inc., Hoffman Estates, IL 60192, USA}
\affil[3]{X-Ray Science Division, Argonne National Laboratory, Lemont, IL 60439, USA}
\affil[4]{Department of Electrical and Computer Engineering, Northwestern University, Evanston, IL 60208, USA}


\maketitle

\begin{abstract}
Photon-counting detectors based on CZT are essential in nuclear medical imaging, particularly for SPECT applications. Although CZT detectors are known for their precise energy resolution, defects within the CZT crystals significantly impact their performance. These defects result in inhomogeneous material properties throughout the bulk of the detector. The present work introduces an efficient computational model that simulates the operation of semiconductor detectors, accounting for the spatial variability of the crystal properties. Our simulator reproduces the charge-induced pulse signals generated after the X/$\gamma$-rays interact with the detector. The performance evaluation of the model shows an RMSE in the signal below 0.70\%. Our simulator can function as a digital twin to accurately replicate the operation of actual detectors. Thus, it can be used to mitigate and compensate for adverse effects arising from crystal impurities.
\vspace{0.2 cm}
\end {abstract}

\vspace{-3mm}


\section{Introduction}
Nuclear medical imaging plays a critical role in modern healthcare, enhancing diagnostic capabilities for a wide range of health conditions. X/$\gamma$-rays detectors are essential components in various medical imaging modalities, including computed tomography, positron emission tomography, and single-photon emission computed tomography (SPECT). These high-energy radiation detectors are often classified as energy integration detectors (EID) or photon-counting detectors (PCD) \cite{detectors-classification1, detectors-classification2, detectors-classification3}.

EIDs were fundamental to the early development of CT diagnostic imaging technologies in the 1970s \cite{first-scintillators1, first-scintillators2}. A typical EID configuration consists of a scintillator coupled with a conventional photodetector. Through the process of luminescence, the scintillator converts X/$\gamma$-rays into low-energy photons (usually in the visible or ultraviolet range), which are then detected by the photodetector \cite{scintillators-low-cost-standard-sensor}. EIDs capture the total energy deposited by multiple incoming photons over a specific time period. As a result, the output signals do not provide information about the particular energy of the incoming photons, leading to the production of single-channel images.

In contrast, PCDs can inherently discriminate the energy of individual photons by generating distinct electrical pulses for each photon-detector interaction, leading to multi-channel images. In particular, CZT detectors have gained relevance due to their high stopping power, which is attributed to the relatively high atomic number of the compound. This feature facilitates the efficient absorption of incoming radiation \cite{CZT_properties_stoppingPower}. In addition, these detectors can operate effectively at room temperature \cite{CZT_properties_roomTemperature_wideSpectralEnergy}, unlike conventional Ge detectors, which require complex cooling systems. CZT detectors can achieve high temporal resolution (on the order of a few nanoseconds) and submillimeter spatial resolution (with pixelated configurations) \cite{zheng2016improving}. 

\begin{figure*}[h!]
\centering
\includegraphics[width=\textwidth]{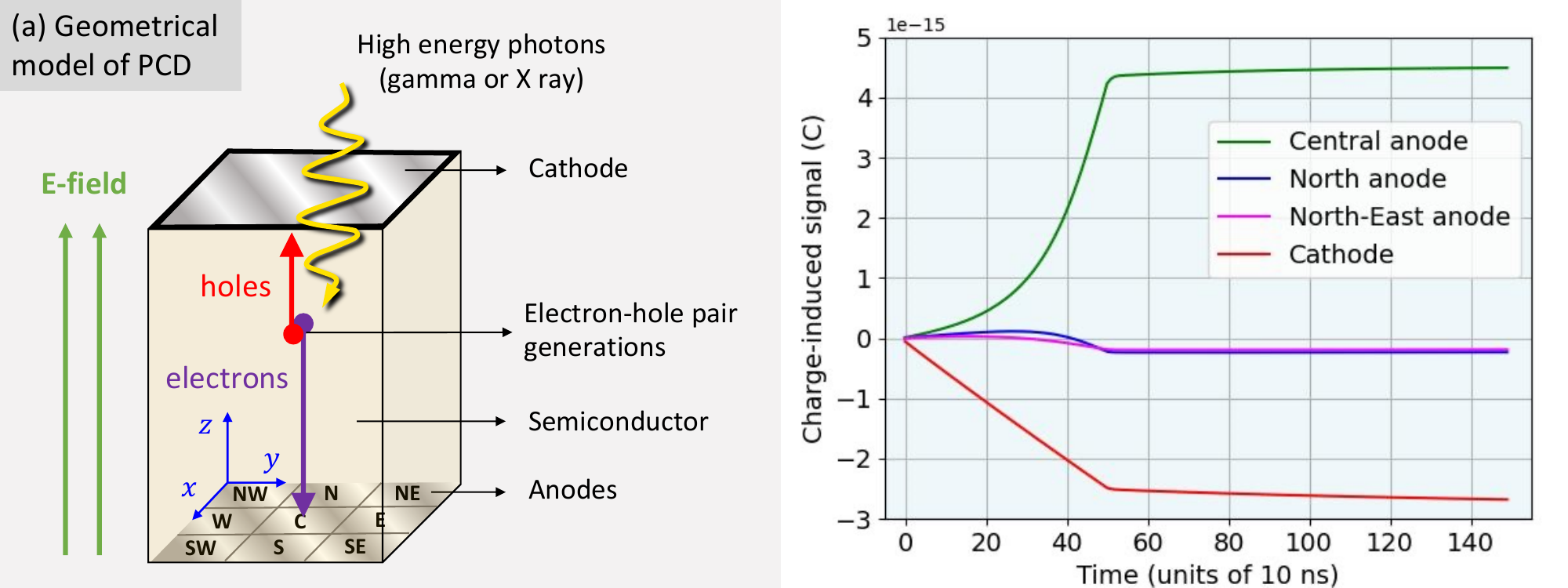}
\caption{(a) Geometrical model of the photon-counting semiconductor detector with the cathode on top and an array of nine anodes (or pixels) at the bottom. (b) Charge-induced signals at different electrodes.}
\label{fig:geometry}
\end{figure*}

It should be noted that single-photon CZT detectors are not suitable for high-photon flux conditions, as in conventional computed tomography. In these cases, the pulse signals from one and the next photons would pile up, preventing an effective distinction between them \cite{PCD_pileup}. These detectors have now become prominent in SPECT applications, which often operate with just a few cps/mm$^2$ \cite{montemont2014studying}. In SPECT imaging, the isotope typically used is $^{99\text{m}}$Tc, which emits around 98. 6\% of $\gamma-$decay at an energy of 140.5 KeV \cite{huh2023simulation}. Taking into account a linear attenuation coefficient of around $\alpha=2.8$ cm$^{-1}$ for CZT at this energy, the Beer-Lambert law \cite{hecht2012optics} tells us that a detector of $z=1$ cm thick would stop around 94\% of the intensity of an incident collimated beam, since the transmitted intensity is $I_\text{T}=I_0 e^{-\alpha z} \approx 0.06 I_0 $. This rough calculation highlights the need for thick detectors in SPECT applications.

A major limitation of these thick CZT detectors is the relatively high concentration of defects throughout the crystal volume \cite{defects1, defects2, defects3, defects4}, which ultimately affects the accurate reconstruction of photon-detector events. The recent literature \cite{table_Srutarshi, ballester2022materials, banerjee2023machine, banerjee2024physics, banerjee2023identifying} has increasingly focused on characterizing the material properties and defects of the detector bulk, with the goal of mitigating and compensating for the adverse effects introduced by the impurities of the crystals. These programs usually employ a \textit{quick} forward model that closely simulates the actual detector. These programs then solve an inverse problem (using either traditional optimization or novel machine learning techniques) and look for the actual material properties and defects that best fit the experimental signals with the computer-generated ones. Therefore, it becomes critical to have a high-fidelity model that accurately captures the complexities of the crystal and realistically simulates detector operations.

Motivated by this problem, we present a novel computational model that simulates the functioning of semiconductor-based PCDs with high efficiency, running in just 50-100 ms on a standard computer. Our model reproduces the charge-induced pulse signals generated when $\gamma-$rays interact with the detector. The simulator takes as inputs: the 3D injection position $r=(x,y,z)$, the energy deposited $E_\text{d}$, and the spatial distribution of the properties and defects of the detector bulk, considering a multilayer model from cathode to anode. The simulator then outputs the charge-induced signals at each electrode. Compared to the expected theoretical results, we found that our signals have an RMSE of less than 0.70\%.

Numerous simulators have been described in the literature \cite{comparison_guerra, simulator_Efield, benoit2009simulation, table_hecht}, each varying in complexity, computational duration, and accuracy. To our knowledge, the present work introduces the first \textit{fast} simulator that incorporates the effects of recombination, trapping, detrapping, and diffusion as spatially dependent multilayer features. Our Python programs are publicly available at this \href{https://drive.google.com/drive/folders/1JXiXKuunksDnAZ4ZNCRDYv-ha0HQ62pu?usp=sharing}{link}.

\section{Simulation framework and assumptions}
\label{sec:PCDs}

\begin{table*}[]
    \centering
    \captionsetup{width=0.8\textwidth} 
    \begin{tabular}{|c|c|c|c|}
        \hline
        \rowcolor{blue1}
        Detector properties & Symbol & Common range & Selected values \\
        \hline
        \rowcolor{blue2}
        E-field strength [V/cm] & $E$ & 500-2000 & 1000 \\
        \hline
        \rowcolor{blue3}
        Mobility [cm$^2$/Vs] & $\mu_\text{e}$ & 1000-1350 & 1000 \\
        \cline{2-4}
        \rowcolor{blue3}
         & $\mu_\text{h}$ & 50-120 & 100 \\
        \hline
        \rowcolor{blue3}
        Recombination lifetime [\textmu s] & $\tau_\text{eR}$ & $2.3-10.0$ & 10 \\
        \cline{2-4}
        \rowcolor{blue3}
         & $\tau_\text{hR}$ & $1.0-2.3$ & 1 \\
        \hline
        \rowcolor{blue2}
        Trapping lifetime [\textmu s] & $\tau_\text{eT}$ & $1-130$ & 10 \\
        \cline{2-4}
        \rowcolor{blue2}
         & $\tau_\text{hT}$ & $0.05-7.00$ & 0.07 \\
        \hline
        \rowcolor{blue3}
        Detrapping lifetime [\textmu s] & $\tau_\text{eD}$ & $0.02-0.40$ & 0.40 \\
        \cline{2-4}
        \rowcolor{blue3}
         & $\tau_\text{hD}$ & $0.01-5.00$ & 0.07 \\
        \hline
    \end{tabular}
    \caption{
    Average material properties of CZT detectors. Typical values reported in the literature for the electric field \cite{E-fields}, the mobility of electrons \cite{table_1, table_2, table_hecht} and holes \cite{table_1, table_2, table_hecht}, the recombination lifetime of electrons \cite{table_2, table_3, table_4} and holes \cite{table_2, table_6}, the trapping lifetime of electrons \cite{table_1, table_3, table_5} and holes \cite{table_1, table_5}, and the detrapping lifetime of electrons \cite{table_7, table_Nature} and holes \cite{table_Miesher, table_Nature}.}
    \label{tab:1}
\end{table*}

Semiconductor PCDs can be constructed in various configurations, including planar, pixelated, or coaxial. In this study, we consider a pixelated geometry, where the cathode and the anodes are arranged on opposite sides, and the anode side is subdivided into an array of small pixels, allowing for higher spatial resolution, as shown in Fig. \ref{fig:geometry}. We will analyze a detector of dimensions $1 \times 1 \times 1$ cm$^3$ and $3 \times 3$ pixels, maintaining a width to height ratio of 1:1, as commonly reported in the literature \cite{E-fields}. In this configuration, incident radiation penetrates through the cathode side\footnote{Given the substantial stopping power of CZT, most of the events will take place near the cathode. The setup described above is advantageous for generating short pulses and, therefore, avoiding the accumulation of consecutive signals. The (slow-moving) holes are quickly collected in the cathode, whereas the (fast-moving) electrons traverse the whole bulk rapidly.} and interacts with the semiconductor material. When $\gamma$-radiation arrives at the CZT crystal, several events occur. The primary mechanisms of photon absorption for the energy ranges used for nuclear medical imaging involve Compton scattering and the photoelectric effect \cite{miesher_model}. Photon absorption initiates the transition of the electrons from the valence band to the conduction band, resulting in the formation of electron-hole (e-h) pairs. Given an ionizing energy of $E_\text{ion} = 4.6$ eV and a single $\gamma-$photon of $E_\text{d}=140.5$ KeV, approximately $P=E_\text{d}/E_\text{ion}=30543$ electron-hole pairs are produced. This corresponds to a charge of $4.86 \times 10^{-15} C$ for both electrons and holes. The free electrons and holes move through the semiconductor under the influence of an operative (external) electric field $E_\text{op}(z)$. While the electron propagates toward the anodes, the hole is directed toward the cathode. Ultimately, the motion of these elementary charges generates charge-induced signals at the nearby electrodes \cite{SR_review}. 

\begin{figure}[htbp]
\includegraphics[width=\columnwidth]{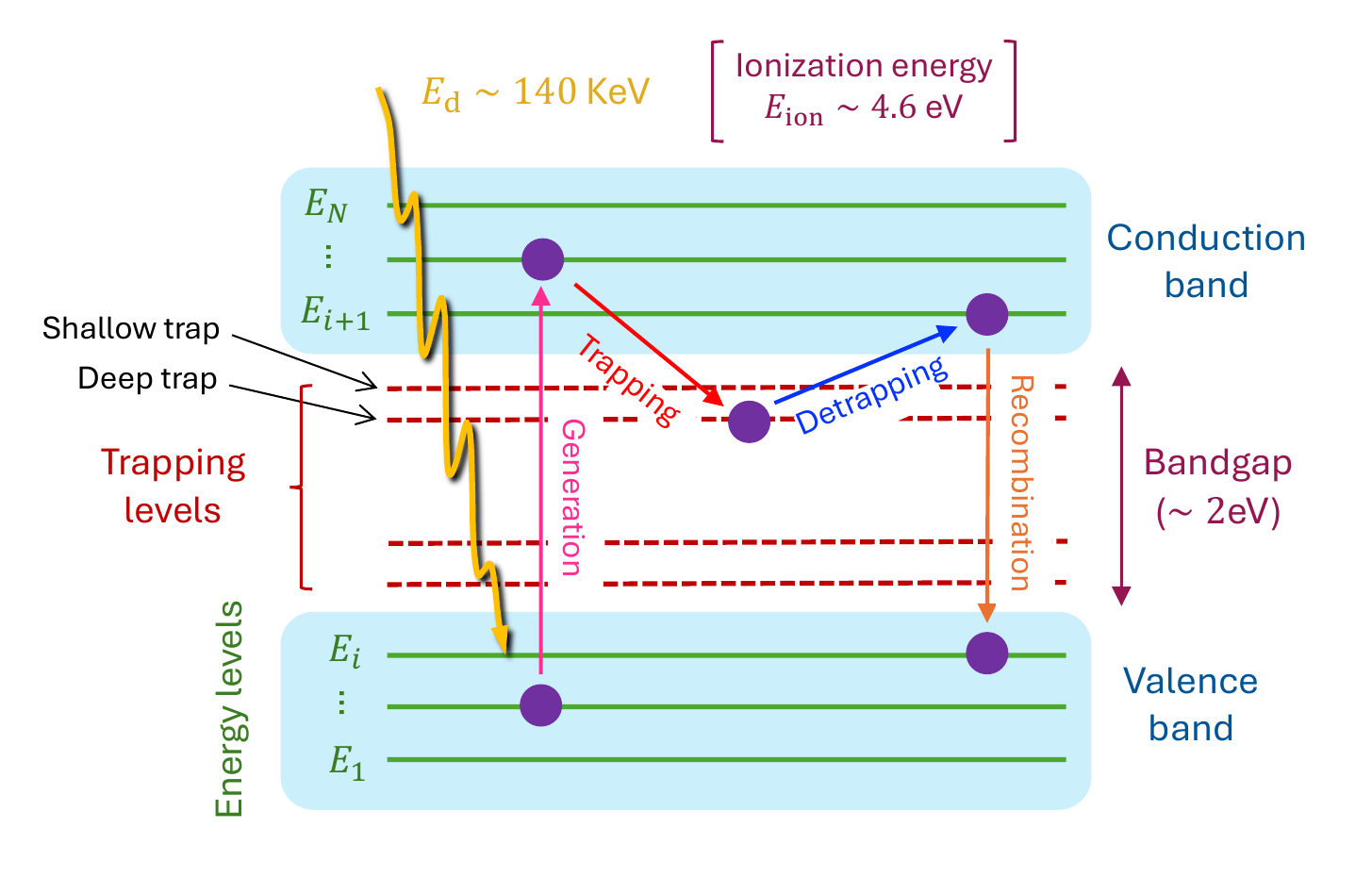}  
\caption{Energy level diagram explaining the electron-hole charge generation, recombination, and trapping-detrapping processes.}
\label{fig:energy_levels}
\end{figure}

The following properties included in our model describe how the charges propagate within the detector: 
\begin{itemize}
    \item \underline{Drift velocity}: The average velocity at which charge carriers move through the semiconductor depends on the external operative field. Being $\mu_\text{e}$ and $\mu_\text{h}$ the mobility of electrons and holes (cm$^2$/Vs), the drift velocities of the charges in the $z-$direction are expressed as $v_\text{e}(z) = -\mu_\text{e}(z) E_\text{op}(z)$ and $v_\text{h}(z) =\mu_\text{h}(z) E_\text{op}(z)$.

    \item \underline{Diffusion}: When charges are generated, they can be represented by a highly localized delta function, $\delta(x_0,y_0,z_0)$. As time progresses, these charges begin to spread outward around a center point, ideally producing a spherically symmetric charge distribution. According to the Einstein-Smoluchowski relation, diffusion in semiconductors depends on the coefficient $ D = \mu k_\text{B} T/q$. Here, $\mu$ represents the charge mobility for electrons or holes, $k_\text{B}$ is the Boltzmann constant, $q$ is the elementary charge, and $T$ denotes the temperature, which is assumed to be 300K (room temperature) in our simulations. We will denote $D_\text{e}$ for electrons and $D_\text{h}$ for holes. It is important to highlight that the diffusion coefficient $D$ is directly proportional to charge mobility and is completely determined by that term (at room temperature). An electron diffusion coefficient of approximately 25 cm$^2$/s is common for CZT crystals.

    \item \underline{Carrier recombination lifetime}: The free charges may undergo recombination when the electrons transition back from the conduction band to the valence band, as seen in Fig. \ref{fig:energy_levels}. The average time before electrons and holes recombine is represented as $\tau_\text{eR}(z)$ and $\tau_\text{hR}(z)$, respectively. To estimate the length that the average charge travels before recombination occurs, one can multiply the operative field by the product of mobility life, $\mu_\text{e} \tau_\text{eR}$ and $\mu_\text{h} \tau_\text{hR}$ (cm$^2$/V) \cite{mobility-lifetime}.

    \item \underline{Carrier trapping and detrapping lifetimes}: Carriers can be trapped at specific (shallow or deep) trapping energy levels \cite{trapping_prettyman, table_Miesher}, as seen in Fig. \ref{fig:energy_levels}. These traps temporarily prevent the charges from contributing to the signal. Eventually, the trapped charges may be released, leading to additional signal production. The trapping lifetimes, $\tau_\text{eT}(z)$ and $\tau_\text{hT}(z)$, and the detrapping lifetimes, $\tau_\text{eD}(z)$ and $\tau_\text{hD}(z)$, are different for each of the energy levels. For simplification, our model employs a single \textit{average} trapping and detrapping lifetime, as done in \cite{table_1, table_Nature}.

\end{itemize}

Table \ref{tab:1} presents a range of typical values for CZT detectors reported in the literature and the specific parameters used in our simulations. Our model considers that such properties have a particular value in each stratified medium ($N = 100$ layers in our simulations) arranged from cathode to anode. The spatial properties will vary significantly depending on the preparation conditions and the specific stoichiometry of the crystal \cite{CZT_properties_noise, mandal2014characterization, kim2015material}, as well as the general condition of the detector. Nevertheless, it is important to recognize that these material properties could, in reality, also vary over time and be influenced by other factors, such as temperature, generation rate, electric field strength, and incident photon energy \cite{table_1, table_2, table_4, table_7, table_hecht_detrap}. 

Additionally, our model neglects the contribution of Coulomb repulsion between the charges and considers that the charges move vertically (see Fig. \ref{fig:geometry}), following the straight field lines of the operative E-field. Therefore, our approach does not include: (i) charge propagation close to interpixel gaps, where $E_\text{op}$ potentially bends in the $x$ and $y$ directions, and (ii) certain polarization effects that also cause deviations in the $x$ or $y$ directions. The latter case occurs when the accumulation of permanently trapped charges within the detector modifies the electric field distribution. As the charge propagation is strictly vertical, the diffusion is considered only through the 1D path and not as a real 3D spherical distribution. By adopting this one-dimensional propagation approximation, we trade off some accuracy for higher efficiency. This assumption is also consistent with the common practices found in the literature \cite{two_assumptions, trapping_prettyman, table_Miesher, table_Srutarshi, table_Nature, ballester2022materials}.

\section{Charge concentration}
\label{sec:charge-concentration}

\begin{figure*}[]
\centering
\includegraphics[width=\textwidth]{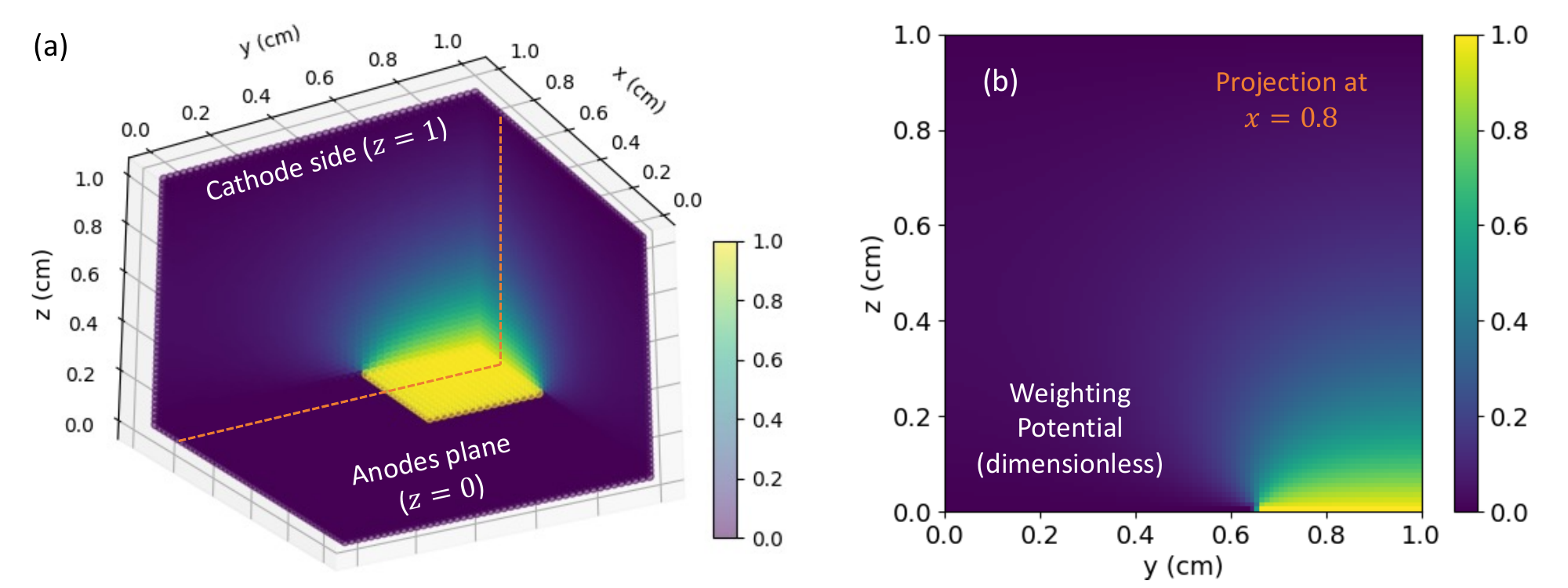}
\caption{(a) 3D view of the weighting potential for a corner anode. (b) Projection of the potential at $x=0.8$ cm.}
\label{fig:weighting_potential}
\end{figure*}

Given an injection at $(x_0,y_0,z_0) \in [0,1]^3$, we now want to calculate the concentrations of electrons and holes in time and space, represented as $n_\text{e}(t,x_0,y_0,z)$ and $n_\text{h}(t,x_0,y_0,z)$, respectively. The propagation of negative charges in a semiconductor can be modeled \cite{book_continuity} using the continuity equation, $\partial_t n_\text{e} - \nabla \cdot (\mu E_\text{op} n_\text{e}) = G - R$. Here, the function $G=\delta(t,x_0,y_0,z_0)$ represents the e-h generation rate at a given location and $R$ the recombination rate. An equivalent formula applies to holes, with the only distinction of a sign change, since the E-field is positive for holes, which move up toward the cathode at $z=1$ (see Fig. 1a). It is then crucial to incorporate the trapping-detrapping effects into this simplistic model so that we can account for the crystal defects. A widely adopted PDE model that accommodates this more intricate scenario is written as follows \cite{trapping_prettyman, table_Miesher, table_Nature}:
\begin{equation}\label{eq:pde}
\partial_t n_\text{e} = \partial_z \big(\mu_\text{e} E_\text{op} n_\text{e}\big) +  D_\text{e} \partial_{zz} n_\text{e}  - \frac{n_\text{e}}{\tau_\text{eR}} - \frac{n_\text{e}}{\tau_\text{eT}} + \frac{n_\text{e}}{\tau_\text{eD}} + \delta
\end{equation}

Here, the function $\tilde{n}_\text{e}(t,x_0, y_0,z)$ denotes the concentration of trapped charges at an average trapping energy level. It can be seen that Eq. \ref{eq:pde} corresponds to the previous continuity equation, now including the loss of charges due to trapping and the addition of charges included during the detrapping process. Additionally, we must include a coupled PDE that explains the evolution of trapped charges over time. A common approach is to considering the trapping centers as \textit{infinite wells} that absorb and expel a fraction of the charges \cite{trapping_prettyman, table_Miesher}:
\begin{equation}\label{eq:well}
\partial_t \tilde{n}_\text{e} = \frac{n_\text{e}}{\tau_\text{eT}} - \frac{n_\text{e}}{\tau_\text{eD}}
\end{equation}

We have used a customized explicit finite difference method (FDM) \cite{numerical_PDEs} to solve the Eqs. \ref{eq:pde} and \ref{eq:well} efficiently (see the Annex for details). Knowing the initial concentrations of electrons and holes, we can then solve these two equations sequentially at each time step. Considering the high temporal resolution of current CZT detectors, a time step of $\Delta t = 10$ ns or less is desired (our simulations use precisely 10 ns). The spatial step $\Delta z$ should be chosen appropriately to maintain the stability of the numerical method. We have set $\Delta z = 0.01$ cm for electrons and $\Delta z = 0.001$ cm for holes. It is important to highlight that Eq. \ref{eq:pde} is overwhelmingly influenced by the advection term (while diffusion plays a minimal role). Thus, it becomes crucial to ensure that the charges do not leap over ($v \Delta t$) more than a voxel ($\Delta z$) at a time. This requirement represents the CFL condition, which is necessary to ensure stability.

\begin{figure*}[h!]
\centering
\includegraphics[width=\textwidth]{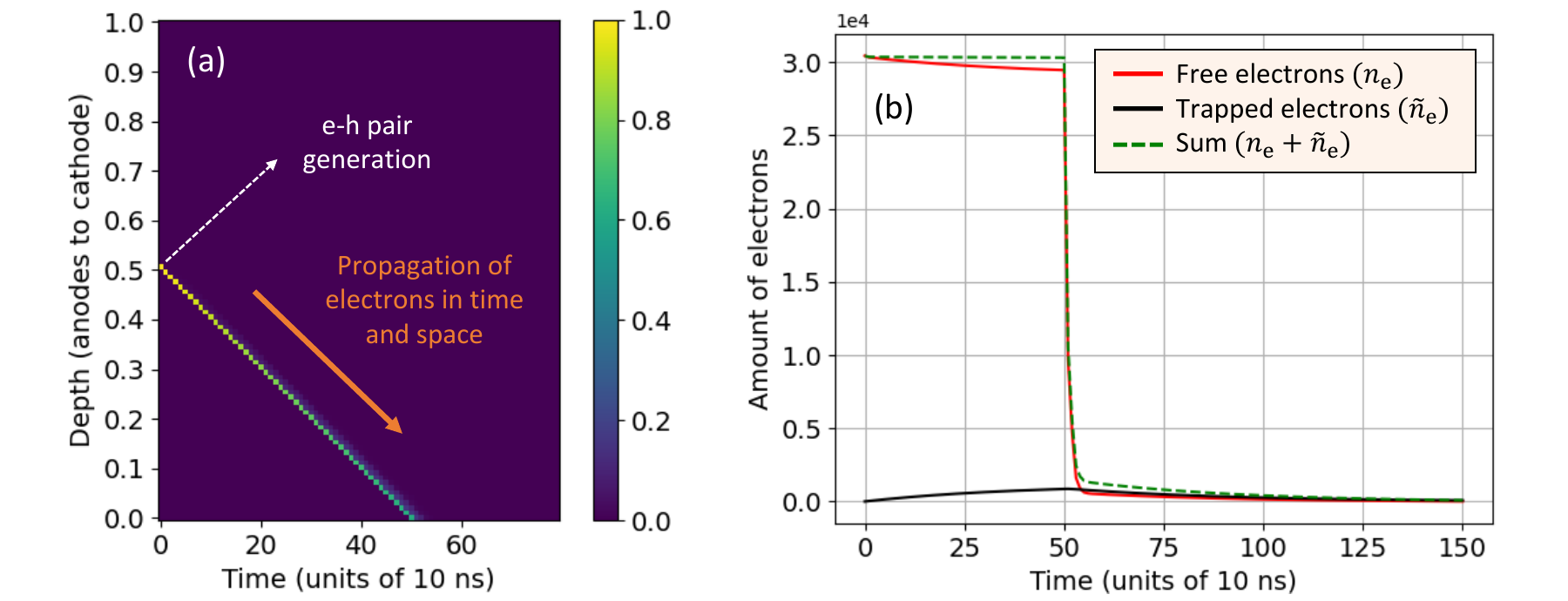}
\caption{(a) Charge concentration of electrons over time and space, $n_\text{e}(t,z)$. The E-field drives the charge carriers from the injection point (voxel 50) to the anode (voxel 100), where they are collected. We employ the material properties described in Table \ref{tab:1}. (b) When electron recombination is neglected ($\tau_\text{eR} \rightarrow \infty$), the amount of charge initially generated should remain constant until collection (at time step 51).}
\label{fig:conservation}
\end{figure*}

\section{Charge-induced signals}
\label{sec:signals}
At each time and position, the free charges with concentrations $n_\text{e}(t,z)$ and $n_\text{h}(t,z)$ generate an electric field $E_\text{g}(t, x,y,z)$ that will eventually reach all the nearby electrodes. As these charges propagate, the field strength will change at the electrode location. Consequently, the electrodes will detect a charge-induced signal $Q(t)$ that varies over time. One can calculate the charged-induced signals at the electrodes using the Gauss law \cite{book_mirror}:
\begin{equation}
    Q(t) = \oiint_S \epsilon \mathbf{E_g}(t) \cdot \mathbf{dS}
    \label{eq:Gauss-law}
\end{equation}

\noindent Here, $S$ is the surface of the electrode of interest and $\epsilon$ the permittivity (the relative permittivity of CZT is $\epsilon_r \approx 11$).  The mirror method \cite{book_mirror} is then typically used to determine $\mathbf{E_g}(t) \cdot \mathbf{dS}$. Following this idea, we would have to compute the double integral from Eq. \ref{eq:Gauss-law} at every time step, making the method computationally expensive. 

The Shockley-Ramo (SR) theorem provides an alternative formulation that is significantly more efficient. This theorem assumes that, at a given time, the drift velocity of the charges is negligible in comparison with the speed of the E-field emanating from the carrier to the electrodes. At any given moment, we can then approach the problem using electrostatic theory \cite{SR_review}. To apply the SR theorem, it is necessary first to determine the ``weighting fields'' associated with each electrode. These fields describe how a charge placed at a specific location influences the signal of the given electrode. 

For each electrode $i \in \{1,2,...,10\}$  (one cathode and nine anodes) with surface $S_i$, we need to solve the 3D Poisson equation subjected to Dirichlet boundary conditions:

\begin{subequations} \label{eq:poisson}
\begin{align}
    & \nabla^2 \varphi (x,y,z) = 0\\
    & \varphi_{|S_i} = 1\\
    & \varphi_{|S_k} = 0, \text{      for  } i \neq k \in \{1,2,...,10\}
\end{align}
\end{subequations}

\noindent The solutions were computed using the Finite Element Method (FEM) in Python using the package \textit{FEniCS} \cite{langtangen2017solving}. Fig. \ref{fig:weighting_potential}a shows the 3D potential for a corner anode, while Fig. \ref{fig:weighting_potential}b shows the cross section at $x=0.8$. We then calculated the weighted electric field $E_w^{(i)}(z)$ using central differences.

Considering now known the charge concentrations and the weighting electric field, the SR theorem provides a simple formula \cite{table_1} to derive the current signal $I$ at each electrode $i$ induced by the electrons (and equivalently, for holes):
\begin{equation}
    I_\text{e}^{(i)}(t) = \int_z \big( q \cdot n_\text{e}(z,t) \big) v_\text{e}(z) E_w(z) dz
    \label{eq:SR}
\end{equation}
The numerical integral becomes a summation in all $N$ voxels. Subsequently, we can numerically integrate the current in time to obtain the charge-induced signals $Q_\text{e}^{(i)}(t)$ at each electrode. Ultimately, the total signal at each electrode $i$ is the sum of the signals induced by electrons and holes, $Q^{(i)} = Q_\text{e}^{(i)} + Q_\text{h}^{(i)}$. Note that electrons will contribute significantly more than holes to anode signals \cite{table_3, table_4}, given their higher velocity.

\section{Propagation and conservation of charges}

We first examine the propagation of charges in both time and space. In this case, we set all voxels with the same material properties, as seen in Table \ref{tab:1}. We focus on a single injection at the center detector, precisely above the central anode, and in the voxel 50 out of 100 (depth z = 0.5 cm). Fig. \ref{fig:conservation}a illustrates the propagation of electrons in time and space. The clear diagonal slope of the charges shows that the electrons move approximately one voxel per unit of time, meaning that the average velocity is about $v_\text{e} = \Delta z/\Delta t = 10^6$ cm/s. This quantity completely matches the expected drift velocity, $v_\text{e} = \mu_\text{e} E_\text{op}$, partially showing the coherence of the model. Fig. \ref{fig:conservation}a also shows a very slight spread of the charge over time as a result of the diffusion effect and a reduction of the amount of charge due to recombination and trapping.

It is now critical to demonstrate numerically the conservation of charges within the system, as it is imposed by the continuity equation when $G - R = 0$. Therefore, when the recombination process is ignored ($1/\tau_\text{R} \approx 0$), the generated charges must remain in the conduction band as free charges or be trapped at certain energy levels. Fig. \ref{fig:conservation}b shows the sum of the calculated concentrations of free and trapped electrons, which remains invariant until the carriers are collected at the electrodes. Given the displacement $d = 0.5$ cm from the injection point to the anode, the exact time required to collect the generated electrons should theoretically be $t = d/v_\text{e} = 50$ $(\times 10$ ns), as observed on the graph\footnote{After the time step 50, it can be seen in Fig. \ref{fig:conservation}b that most of the electrons are collected. However, there are still some electrons left behind due to diffusion that will slowly arrive at the anodes over time. Diffusion makes the collection of charges smoother.}.

\section{Comparison: Ramo and Hecht models}

\begin{figure}[htbp]
\includegraphics[width=\columnwidth]{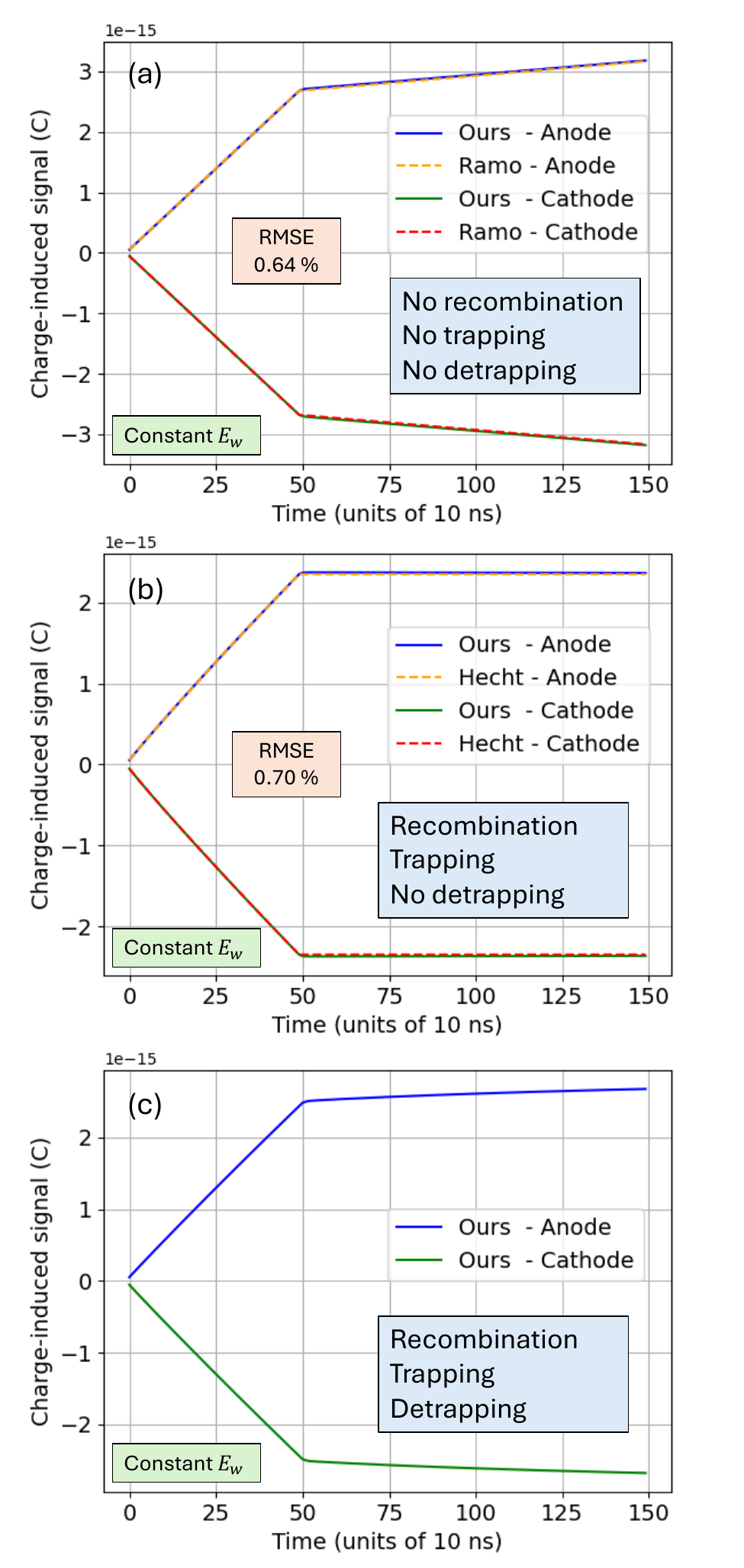}  
\caption{Planar detector with one cathode on the top and one anode on the bottom. Linear operative field and constant weighting field. (a) Ramo model, which does not consider recombination, trapping or detrapping. (b) Hecht model, which considers recombination and trapping. (c) Our model based on the FDM, which considers recombination, trapping and detrapping.}
\label{fig:Ramo_Hecht}
\end{figure}

To further assess the accuracy of our model, we consider a simple scenario where Eq. \ref{eq:pde} has an analytical solution. If we neglect the detrapping process ($1/\tau_\text{D} \approx 0$) and charge diffusion ($D \approx 0$), then the two differential equations in Eq. \ref{eq:pde} can be easily decoupled and solved independently. The charge concentration (for electrons or holes) then becomes:
\begin{equation}
    n(t,z) = n\big(0, z_0 \pm v t \big) e^{-t \big( 1/\tau_R + 1/\tau_T \big)}
    \label{eq:analytical}
\end{equation}

Observe that the analytical solution in Eq. \ref{eq:analytical} is a product of two terms. The first term accounts for charge transport until collection at the electrode, where the sign of the velocity depends is positive for holes and negative for electrons. The second term describes the exponential reduction in charge due to trapping and recombination. Having $n(t,z)$ calculated from Eq. \ref{eq:analytical}, one can easily determine the concentration of the trapped charge as $\tilde{n}(t,z) = 1/\tau_\text{T} \int n(t,z) dt$. 

In this section, we will also consider for mere comparison reasons a planar (not pixelated) detector configuration, with an ideally infinitely extended cathode plane on the north side ($z=1$) and another extended anode on the south side ($z=0$). In that case, Eqs. \ref{eq:poisson} have analytical solutions. The weighted potentials are linear and given by:
\begin{subequations}
\begin{align}
    & P_\text{w}^\text{Cathode} = \frac{z}{\text{Depth}} = \frac{v t + z_0}{\text{Depth}}
\label{eq:p1} \\
    & P_\text{w}^\text{Anode} = \frac{\text{Depth} - z}{\text{Depth}} = \frac{\text{Depth} - (v t + z_0)}{\text{Depth}}
 \label{eq:p2}
\end{align}
\end{subequations}

\noindent In our simulations, the depth is 1 cm and the corresponding weighted fields are clearly constant ($-1$ cm$^{-1}$ for the cathode and 1 cm$^{-1}$ for the anode). Note that both the Ramo \cite{SR_ramo} and Hecht models \cite{table_hecht} assume this simplistic detector geometry. 

In particular, the Ramo equation only considers the transport of a charge, not including fluctuations in the amount of charge. The SR theorem states that $I(t) = \rho(t) v E_\text{w}$ for a localized charge density $\rho(t)$, which is a particular case of the more general Eq. \ref{eq:SR}. Taking into account then only the first factor of Eq. \ref{eq:analytical}, the charge density of electrons and holes simply becomes $\rho(t) = q \cdot n \big(0, z_0 \pm v t \big)$, until collection. We can find the charge-induced signal at one of the electrodes $i \in \{1, 2\}$ using the fundamental theorem of calculus:
\begin{equation}
    Q^{(i)}_\text{Ramo}(t) = \int_0^t I^{(i)}(t) dt = q \big(P^{(i)}_w(t) - P^{(i)}_w(0) \big)
 \label{eq:Ramo}
\end{equation}

\noindent Fig. \ref{fig:Ramo_Hecht}a shows the linear contribution of electrons since their generation (at the central depth) to their collection at time $t = 50$ $(\times 10)$ ns. The holes, which move about ten times slower (see Table \ref{tab:1}), do not reach the cathode during the analyzed time period. The slight contributions of holes to the signals can be seen after the electrons have been collected, as there is still a shallow slope. As the weighting fields are constant and of opposite sign for the cathode and anode, the signals are expected to be completely symmetric on the ordinary axis, as observed in all subfigures of Fig. \ref{fig:Ramo_Hecht}.

In contrast, the Hecht equation \cite{table_hecht} also includes trapping and recombination. This more advanced model uses a variable $L = \frac{v}{1/\tau_\text{R} + 1/\tau_\text{T}}$, which accounts for the distance traveled by the average charge (for electrons or holes) before it is recombined or trapped. The signal can be calculated as follows
\begin{equation}\label{eq:hecht}
\begin{aligned}
& Q^{(i)}_\text{Hecht}(t) = \int_0^t \bigg\{ q \, n \left(0, z_0 \pm v \cdot t \right) e^{-t \big( 1/\tau_\text{R} + 1/\tau_\text{R} \big)} \\
& v \, E_\text{w}^{(i)} \bigg\} dt  = q \cdot L \cdot E_\text{w}^{(i)} \left( 1 - e^{-t \big( 1/\tau_\text{R} + 1/\tau_\text{T} \big)} \right)
\end{aligned}
\end{equation}

Fig. \ref{fig:Ramo_Hecht}b displays the signals calculated with the Hecht model, As the holes have a relatively low lifetime (see Table \ref{tab:1}), they are quickly trapped. Therefore, we no longer observe their contribution after the electrons are trapped in time step 50, and the signal plateaus after that. Fig. \ref{fig:Ramo_Hecht}c corresponds to the signals generated with our model when now we also include the detrapping process. In the latter case, one can note that the detrapping makes the contributions of holes more relevant since they now get to contribute after being detrapped.

Observe that both Figs. \ref{fig:Ramo_Hecht}a and \ref{fig:Ramo_Hecht}b show that our numerical model, when running on this simplistic planar detector configuration, generates signals that completely superimpose those calculated with the theoretical Ramo and Hecht equations. The RMSE (for both the cathode and the anode signals, given the symmetry) is 0.64\% for the Ramo model and 0.70\% for the Hecht model, respectively. These small discrepancies can be traced back to discretization errors that appear when solving the Eq. \ref{eq:pde} with the (efficient) first-order finite differences. These are well-documented numerical dispersion artifacts that depend on the charge velocity and slightly distort the shape of the concentration solution over time \cite{strikwerda2004finite}.

\section{Small pixel effect}

The charge-induced signals presented in Fig. \ref{fig:geometry}b were relatively sharp, steep, and with fast nonlinear growth. These signals corresponded to a pixelated detector where the ratio between pixel size and depth was $R = 0.33$. However, when considering a planar detector with an infinitely extended cathode at the top and a pixel at the bottom ($R \rightarrow \infty$), Fig. \ref{fig:Ramo_Hecht} displayed long and slowly growing linear signals. 

\begin{figure}[h!]
\centering
\includegraphics[width=\columnwidth]{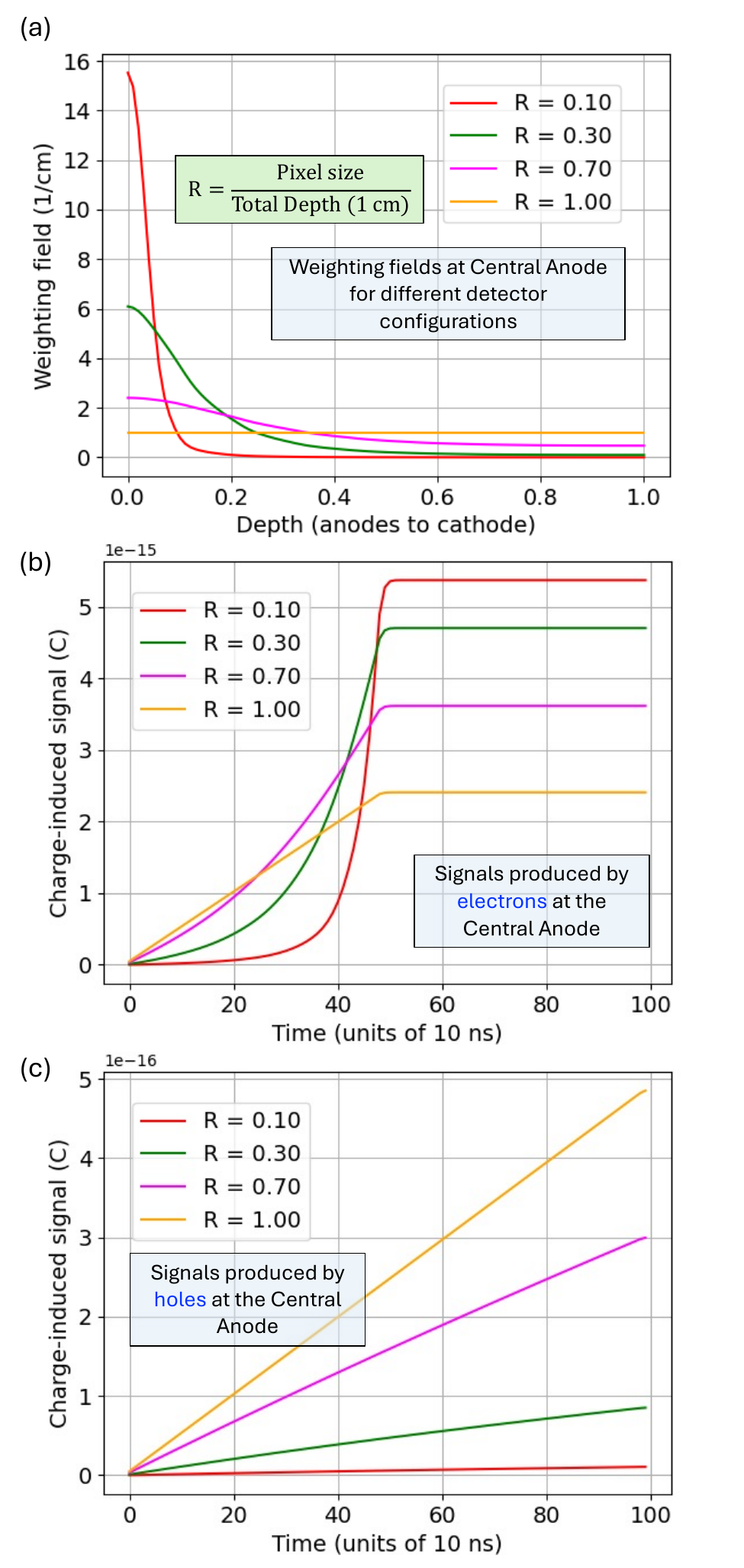}
\caption{(a) Weighting field for the central anode at different depths and fix $(x_0, y_0) = (0.5, 0.5)$ cm. We see the resulting fields different pixel sizes and fix depth of 1 cm. After an event at  $(x_0, y_0, z_0) = (0.5, 0.5, 0.5)$ cm,  we separately show the charge-induced signal at the central anode produced by (b) only the electrons and (c) only the holes.}
\label{fig:pixel_size}
\end{figure}

Barrett et al. \cite{barrett1995charge} examined how pixel size influences signal generation in pixelated detectors. They found that for a detector with a pixel size $\epsilon$, the effective range of the weighting field for a given electrode is approximately limited to a depth $\epsilon$. Fig. \ref{fig:pixel_size}a shows the weighting fields for the central anode for different detector configurations ($R = 0.1, 0.3, 0.7, 1.0$). As the pixel size decreases, we see that the weighting field reaches higher values near the anode ($z = 0$) and decays faster. 

The SR theorem then explains how signal production (see Eq. \ref{eq:SR}) is affected by that weighting field. We now take into account only the free propagation of charges (without recombination, trapping, or detrapping) after injection at position $(x_0, y_0, z_0) = (0.5, 0.5, 0.5)$ cm. Fig. \ref{fig:pixel_size}b shows how the signals generated by the electrons vary for different pixel sizes. The smaller the pixel, the sharper and higher the charge-induced signal, enabling short pulses. 

Given the proposed detector configuration, most pairs will be produced in practice near the cathode. As the holes move towards the cathode, chances are high that they will not pass through the effective range of the weighting potential of the anodes. They then contribute less to the anode signal as the pixel becomes smaller, as seen in Fig. \ref{fig:pixel_size}c. Following these arguments, some simulators (such as those reported in \cite{zumbiehl2001modelling}) neglect hole displacement for simplification. However, it should be emphasized that the contribution to the hole will ultimately depend on the initial depth of injection and is significant for the cathode signal.

\section{Noise sensitivity analysis}
\label{sec:noise}

\begin{figure*}[]
\centering
\includegraphics[width=\textwidth]{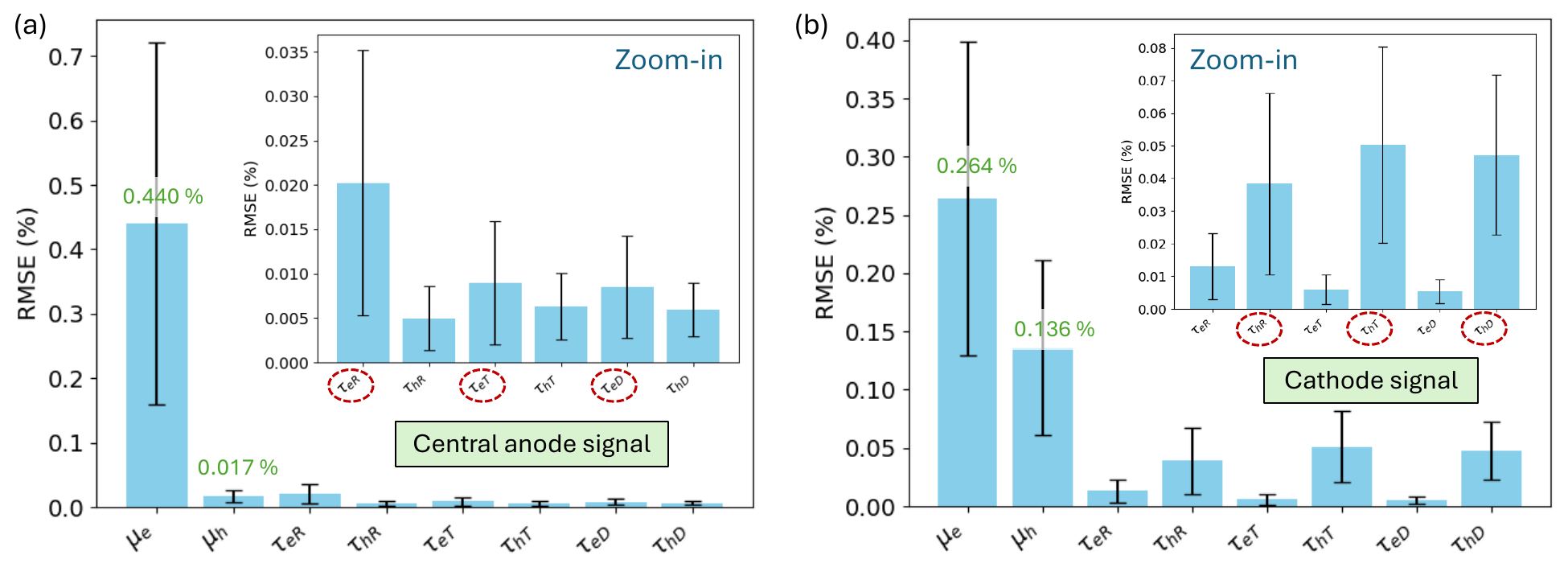}
\caption{Means and standard deviations between a ground-truth signal (generated with parameters from Table \ref{tab:1}) and signals generated with noisy inhomogeneous properties. We see how the introduction of noise at certain parameters (such as $\mu_\text{e}$) strongly distorts the signals, whereas the addition of noise to the lifetime parameters leads to less signal deformation. A spatial Gaussian noise with coefficient of variation of 6\% was added to the voxels of each parameter. (a) Errors in the central anode signal. (b) Error in cathode signals. Both graphs have insets with zoom-in to the lifetime parameters. }
\label{fig:noise}
\end{figure*}

We will now simulate the inhomogeneities and defects inherent in the crystal structure and see how much they affect the signals. In particular, we want to determine how much the addition of noise to specific parameters deforms the signals. The designed experiment introduces Gaussian noise across different voxels for a given material property. This additive noise has a coefficient of variation of 6\%, in agreement with a reasonable uncertainty level reported in the literature \cite{std1}. 

We calculate the RMSE between a ground-truth signal (generated with homogeneous material properties) and the signal produced with a spatially dependent inhomogeneous property. To account for the stochastic nature of the defects, we consider 50 sample signals that were produced when adding different random noises. Figs. \ref{fig:noise}a and \ref{fig:noise}b illustrate bar graphs with the average RMSE and standard deviation between the baseline signals and the noisy ones for the central anode and cathode signals, respectively. 

Because our pixelated configuration takes advantage of the small-pixel effect, we clearly see that most of the signal deformation is due to fluctuations in the electron drift. In particular, when adding noise to electron mobility, we find an RMSE of $0.44 \pm 0.28$\% for the central anode signal (Fig. \ref{fig:noise}a) and $0.26 \pm 0.13$\% for the cathode signal (Fig. \ref{fig:noise}b). We also see that the introduction of noise into the mobility of the holes leads to an RMSE of $0.02 \pm 0.01$\% at the central anode (Fig. \ref{fig:noise}a) and $0.14 \pm 0.08$\% at the cathode (Fig. \ref{fig:noise}b). Therefore, the contribution of holes (particularly at the cathode location) is still significant and should not be overlooked.

The relatively large standard deviations shown in Figs. \ref{fig:noise}a and \ref{fig:noise}b reveal the importance of the particular spatial distribution of the noise. If a high variance is found at the voxels where the weighting potential is high (e.g., near the central anode), the signal becomes more distorted. When the zoomed-in windows from the figures are examined, it is evident that electron lifetimes (for recombination, trapping, and detrapping) predominantly affect the central anode signal, whereas hole lifetimes have a greater impact on the cathode signals, as expected. It is important to highlight that the spatial variance of both the electron and hole lifetimes impacts the signals to the same extent, which reiterates the need to include the hole properties in accurate state-of-the-art simulators.

\section{Conclusions}
\label{sec:discussions}

This research presents a model that simulates the operations of photon-counting semiconductor detectors at high speed. Our simulator considered a pixelated CZT detector and incorporates the effects of recombination, trapping, detrapping, and diffusion as spatially dependent properties. Our program then takes the photon-detector interactions as input and outputs the charge-induced signals at each electrode. 

We have evaluated the performance of our method with respect to ground-truth signals (calculated with the Ramo and Hecht models). We found that our signals presented an RMSE of 0.64\% and 0.70\%, respectively. We also show how the uncertainty of material properties and defects influences the resulting pulse signals. In particular, we introduced a realistic level of additive white noise in the parameters, with a coefficient of variation of 6\%. The results indicate that spatial variations in electron mobility have the most significant effects on signal production. Despite the fact that holes have a low contribution to the signals due to the small-pixel effect, including them in the models is essential for achieving accurate results.

Our simulator can be used to solve inverse problems in the field of detector characterization and event reconstructions. In other words, it can be employed to fit simulated signals to real data in order to extract information about the event (position and energy) or the material properties. Advanced non-convex optimization algorithms and machine learning techniques are particularly useful when solving such complicated problems.

We identify two future improvements to our PCD simulator. First, the inclusion of slight bends of the operative field in the $(x,y)$ directions, which might occur due to the interpixel gap or polarization effects. Second, the addition of the Coulomb repulsion to the model, which further contributes to the spread of the charges over time.

\vspace{0.3 cm}

\textbf{Annex:} 
We present here the customized explicit FDM scheme used to solve Eq. \ref{eq:pde} and Eq. \ref{eq:well}. We discretize the equations for holes (moving in the positive $z-$direction towards the cathode) as follows:
\begin{flalign}
\frac{n^{t+1}_z - n^t_z}{\delta t} = & \frac{v_{z+1} n^t_{z+1} - v_z n^t_z}{\delta z} + &\\
D_\text{h} \frac{n^t_{z+1} - 2 n^t_z + n^t_{z-1}}{\delta z^2} &- \frac{n^t_{z+1}}{\tau_\text{hR}} - \frac{n^t_{z+1}}{\tau_\text{hT}} + \frac{\tilde{n}^t_{z+1}}{\tau_\text{hD}} + \delta \notag &\\
\frac{\tilde{n}^{t+1}_z - \tilde{n}^t_z}{\delta t} &=  \frac{n^t_z}{\tau_\text{hT}} - \frac{\tilde{n}^t_z}{\tau_\text{hD}}&
\end{flalign}

For electrons, one just needs to revert the direction of propagation. We then expressed $n^{t+1}_{z}$ and $\tilde{n}^{t+1}_{z}$ as functions of the free and trapped hole concentrations at the previous time step. Taking into account $n_\text{h}(0,z) = \delta(x_0,y_0,z-z_0)$ and $\tilde{n}_\text{h}(0,z) = 0$, we can efficiently solve these discrete equations sequentially (at each time step, first getting $n$ and then $\tilde{n}$).

\vspace{0.5 cm}

\noindent \textbf{Acknowledgments:} The authors would like to express their gratitude to Miesher Rodrigues for their fruitful comments and discussions. 

\noindent \textbf{Funding:} This work was supported by Siemens Medical Solutions USA, Inc. within the scope of a research agreement.

\noindent \textbf{Conflict of interest:} J.K., F.M., and A.H.V. are affiliated with Siemens Medical Solutions USA, Inc.

\section*{Bibliography}
\bibliography{refs}

\end{document}